\documentclass[prb,aps,preprint,eqsecnum]{revtex4}
\usepackage[centertags]{amsmath}
\usepackage{amssymb}
\usepackage{amsthm}
\usepackage{graphicx,subfigure}
\usepackage{epsfig}
\begin{document}

\title{Characterizing dynamical transitions in bistable system using
non-equilibrium measurement of work}

\author{Pulak Kumar Ghosh
and Deb Shankar Ray{\footnote {e-mail address:
pcdsr@mahendra.iacs.res.in}}}

\affiliation{Indian Association for the Cultivation of Science,
Jadavpur, Kolkata 700 032, India}

\noindent
\begin{abstract}
We show how Jarzynski relation can be exploited to analyze the
nature of order-disorder and a bifurcation type dynamical transition
in terms of a response function derived on the basis of work
distribution over non-equilibrium paths between two thermalized
states. The validity of the response function extends over linear as
well as nonlinear regime and far from equilibrium situations.
\end{abstract}

 \maketitle
\section{Introduction} Advancement of micromanipulation techniques
has opened up the new possibility of estimating equilibrium
thermodynamic quantities from non-equilibrium measurements in
recent
years\cite{jar1,jar2,crooks,wang,Carbe,Liphardt,hummer,bussi,hatano,speck,rito}.
The experiments with pulling forces of piconewton magnitude on RNA
molecules\cite{Liphardt,hummer} and dragging colloidal particles in
a fluid\cite{wang,Carbe} have made it possible to measure the
probability distribution of work exerted on a system. Since the
relaxation time of the system concerned in these experiments is too
long compared to the time scale over which the non-equilibrium
measurements are made, work fluctuations over non-equilibrium
irreversible paths play an important role in estimating the free
energy difference between the thermalized initial and final states
of the system. This is reflected in celebrated Jarzynski
relation\cite{jar1,jar2} which relates the work distribution and
the Helmholtz free energy difference $\Delta F$ as
\begin{eqnarray}
e^{-\beta \Delta F} = \langle e^{-\beta W}\rangle
\end{eqnarray}
where the averaging on the exponential function of the work variable
$W$ has been carried out with work distribution $P(W)$, $\beta$
being $1/k_BT$ with $k_B$ and $T$ denoting the Boltzmann constant
and temperature, respectively. Central to these studies are several
results, e. g., the so-called fluctuation theorems expressing
transient violation of second law of thermodynamics where the
systems in question are driven arbitrarily far from equilibrium.

The probability distribution of work $P(W,t)$ done on a system by
manipulating an external agency or force is followed over
irreversible paths for a given protocol. As free energy is the key
quantity for carrying thermodynamic information and for description
of transition between different states, it is apparent that, by
virtue of Jarzynski equality\cite{jar1,jar2} it is possible to probe
 the signature of dynamical transition in the
behaviour of work distribution. The problem is non-trivial since
for a nonequilibrium system, in general, free energy functional can
not be defined. This, in consequence, raises difficulty in
constructing a response function directly in terms of free energy
and its derivatives with respect to suitable parameters of the
system in the spirit of what is done in equilibrium phase
transition. This difficulty, however, can be overcome with the help
of Jarzynski equality by probing work fluctuations on the system
even far from equilibrium. The object of the present paper is to
examine this issue and to look for the response function which is
characteristic of the dynamical system under study and is derivable
from work distribution. In what follows we use a model potential to
explore the order-disorder transition as well as a transition
arising out of bifurcation of the response function in terms of the
time evolution of work distribution function. The dynamics is
described by a Langevin equation which appropriately takes care of
the variation of the potential parameter in inducing transition
from an unimodal to bimodal character\cite{pkg2,ff5}. The
implication of the results are elaborated.

\section{The model and the equilibrium description}
Consider a Brownian particle in a thermal bath at temperature $T$
and subjected to an external potential force. The governing
Langevin equation is given by
\begin{subequations}
\begin{eqnarray}
\dot{x}&=&p\label{1.1a}
\\
 \dot{p}&=&-\gamma p
-V'(x,t)+\Gamma(t)\label{1.1b}
\end{eqnarray}
\end{subequations}
where $V(x,t)$  is the external potential  and $\gamma$ is the
dissipation constant. $x$ and $p$ denote the coordinate and the
momentum of the particle, respectively. Thermal fluctuations
$\Gamma(t)$ of the bath are modeled by Gaussian, zero mean  and
delta correlated noise
\begin{subequations}
\begin{eqnarray}
\langle \Gamma(t)\rangle&=&0\\\label{1.2a}
 \langle
\Gamma(t)\Gamma(t')\rangle&=&2 D\delta(t-t')\label{1.2b}
\end{eqnarray}
\end{subequations}
where $D=\gamma k_B T$ is the strength of the noise. We make use of
potential $V(x,t)$ of the following form
\begin{eqnarray}
V(x,t)=-\frac{1}{2}\left\{a- \lambda(t)a_0\right\}x^2
+\frac{1}{4}bx^4 \label{1.3}
\end{eqnarray}
where $a, \;a_0, \; b$ are the potential parameters and $\lambda(t)$
is a switching parameter. We first allow the system to reach an
equilibrium with the heat bath at temperature $T$ and then switch on
the parameter, infinitely slowly from an initial state $\lambda=0$
to a final state $\lambda=1$. The model has been explored in the
literature under diverse conditions $^{13,14}$. At the initial state
the potential is bistable with two minima at $x=\pm \sqrt{{a/b}}$
and one maximum at $x=0$. The corresponding distribution function is
bimodal in position coordinate($x$)
\begin{eqnarray}
P_0(x,p)\sim \xi(p)\exp{\left[\frac{\;1/2 a x^2-1/4\; b
x^4}{D}\right]}\label{1.4}
\end{eqnarray}
where $\xi(p)=\exp{(-p^2/2K_BT)}$At the final state,  $\lambda=1$,
for $a
> a_0$ the potential is also bistable with two shifted minima at
$x=\pm\sqrt{{(a-a_0)/b}}$ and one maximum at $x=0$. For $a=a_0$, the
potential contains only $x^4$ term and the corresponding
distribution function is unimodal as follows;
\begin{eqnarray}
P_f(x,p)\sim \xi(p)\exp{\left[\frac{-1/4\; b
x^4}{D}\right]}\label{1.5}
\end{eqnarray}
So the switching process induces a transition of the distribution
function from  bimodal($\lambda=0$) to unimodal one($\lambda=1$).
This is depicted in Fig.1(a,b). Fig.1(a) also includes the case for
$a < a_0$ for $\lambda = 1$. The potential well is steeper compared
to previous case ($V(x) =  b x^4$). The corresponding probability
distribution is also shown in Fig.1(b).

\begin{figure} [htp]
\centering \includegraphics[width=16cm,angle=0,clip]{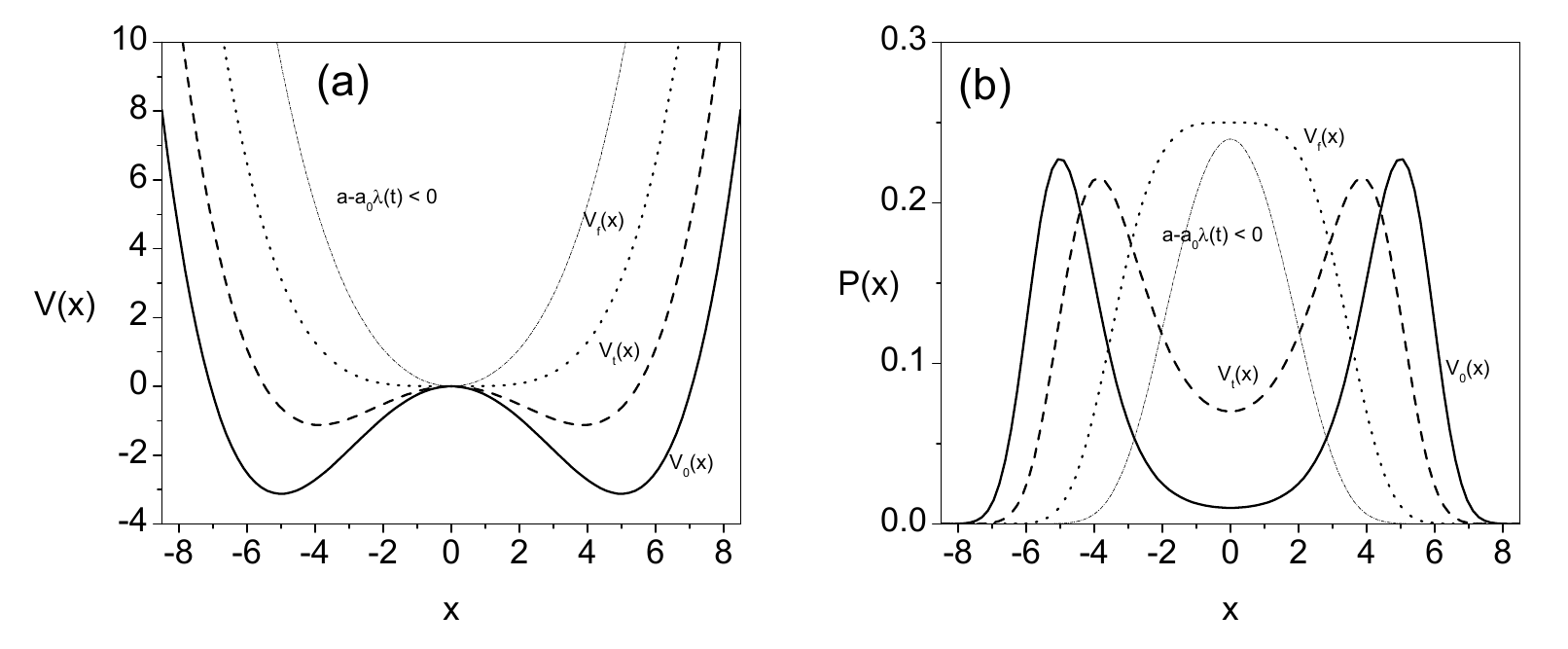}
\caption[]{A schematic illustration for switching of (a) the
potential energy for different values of the parameter  $\lambda$
around the transition point for $(a-\lambda (t)a_0)=a $
corresponding to $V_0$ ($\lambda=0$, solid line), $(a-\lambda
(t)a_0)=a/2$ corresponding to $V_t$ ($\lambda=1/2, $, dashed line,
$(a-\lambda (t)a_0)=0$ corresponding to $V_f$ ($\lambda=1$, dotted
line), and dash-dot line present the case $(a-\lambda (t)a_0)<0$ (b)
distribution function for the corresponding cases. In all the
illustrations $ a=a_{0} $. }
\end{figure}
As both the initial and final states are in equilibrium the free
energies of these states are given by
\begin{eqnarray}
F_0=-k_B T \ln{Q_0}\;;\;\;\;\;\;\;\;\;\;\;\;F_f=-k_B T
\ln{Q_f}\nonumber
\end{eqnarray}
respectively, where $Q$ denotes the partition function. These are
given by the following expressions
\begin{eqnarray}
Q_0&=&\int_{-\infty}^{+\infty}dp \; dx \;
\exp{\left[-\frac{(1/2)\;p^2-(1/2)\; a x^2+(1/4)\; b
x^4}{k_B T}\right]}\label{1.6}\\
Q_f&=&\int_{-\infty}^{+\infty}dp \; dx \;
\exp{\left[-\frac{(1/2)\;p^2-(1/2)\; (a-a_0) x^2+(1/4)\; b x^4}{k_B
T}\right]} \label{1.7}
\end{eqnarray}
respectively. By numerical integration one can easily find out the
change in free energy for arbitrary values of $a_0$.
\begin{eqnarray}
 \Delta F= F_f-F_0=-k_B T \ln{\left(\frac{Q_f}{Q_0}\right)}\label{1.7a}
\end{eqnarray}
The linear coefficient $a_0$ serves here as the constant parameter
of the system. $a_0$ governs the nature of "phase" as a region of
space in which the free energy function is analytical and
continuous. The dynamical transition is associated with the crossing
of boundary between the two regions. To understand the nature of
transition it is therefore worthwhile to look for any discontinuity
or irregularity of the first and second derivatives with respect to
the control parameter at the boundary. With this in mind we, in Fig
2(a), present the variation of free energy change as a function of
$a_0$ at different temperature. In the spirit of traditional way of
deriving thermodynamic information and characterizing equilibrium
phase transition, one may now define an extensive variable, an order
parameter type quantity as, $-\left(\frac{\partial \Delta
F}{\partial a_0}\right)_T$ analogous to magnetization $M$ associated
with the free energy of a magnetic system as
$M=-\left(\frac{\partial F}{\partial H}\right)_T$ (therefore role of
magnetic field $H$ is played by $a_0$). Pushing the analogy a bit
further we define a response function as a second order derivative
with respect to $a_0$ as
\begin{eqnarray}
\Psi=-\frac{\partial^2 \Delta F}{\partial a_0^2} \label{1.7b}
\end{eqnarray}
\begin{figure}[htp]
\centering\includegraphics[width=9cm,angle=0,clip]{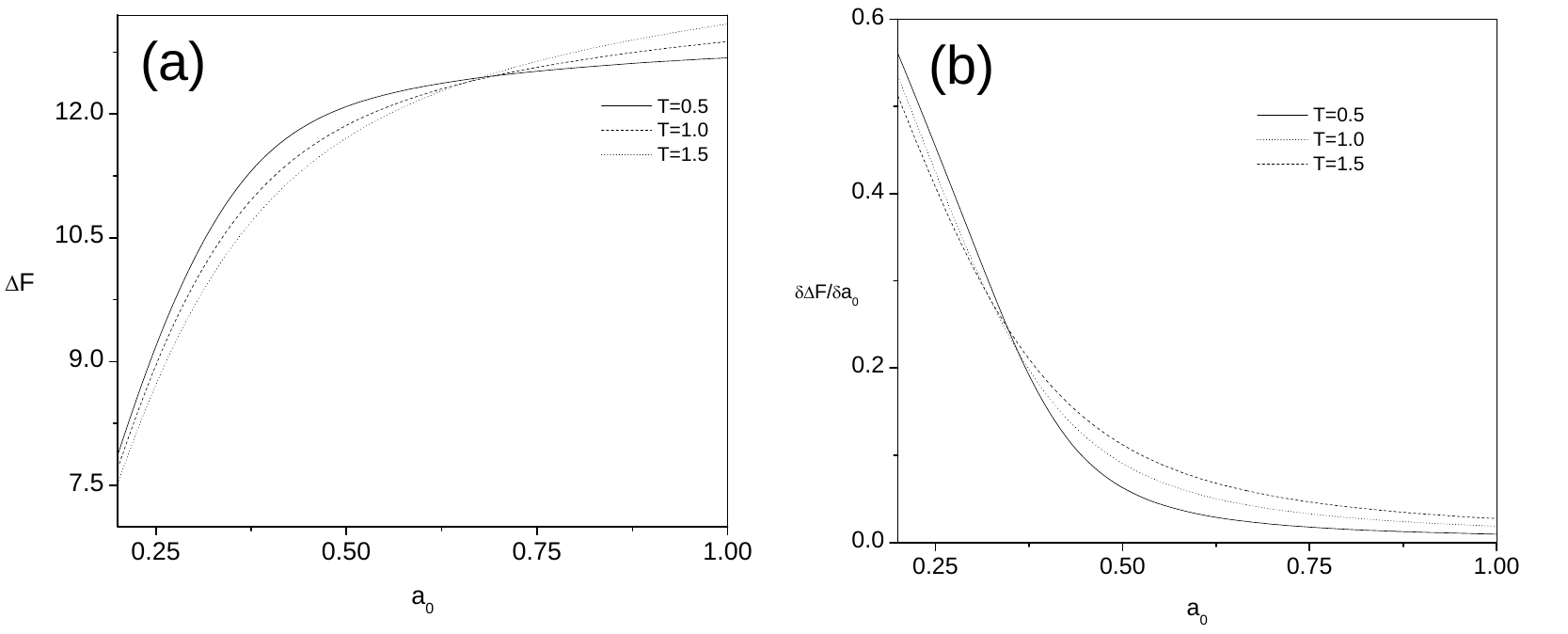}
\caption[]{ (a) Variation of free energy change with $a_0$ at
different temperature, (b)variation of $\frac{\partial\Delta
F}{\partial a_0}$ with $a_0$ at different temperature, for the
parameter set $a=0.5,\;b=0.005,\;K_B=1$.}
\end{figure}
Clearly the above expression is analogous to the magnetic
susceptibility which measures the variation of magnetization due to
the change of the external field. The magnetic susceptibility is
related to the second order derivative of free energy as
$\chi=-\left(\frac{\partial^2  F}{\partial H^2} \right)$. In Fig.2
(b) we depict the first order derivative of free energy change with
respect to $a_0$ at different temperature. While no discontinuity or
singularity is observed in the curve, the behaviour of response
function (2.9) as shown in Fig.3 is markedly different.
\begin{figure}[htp]
\centering\includegraphics[width=8cm,angle=0,clip]{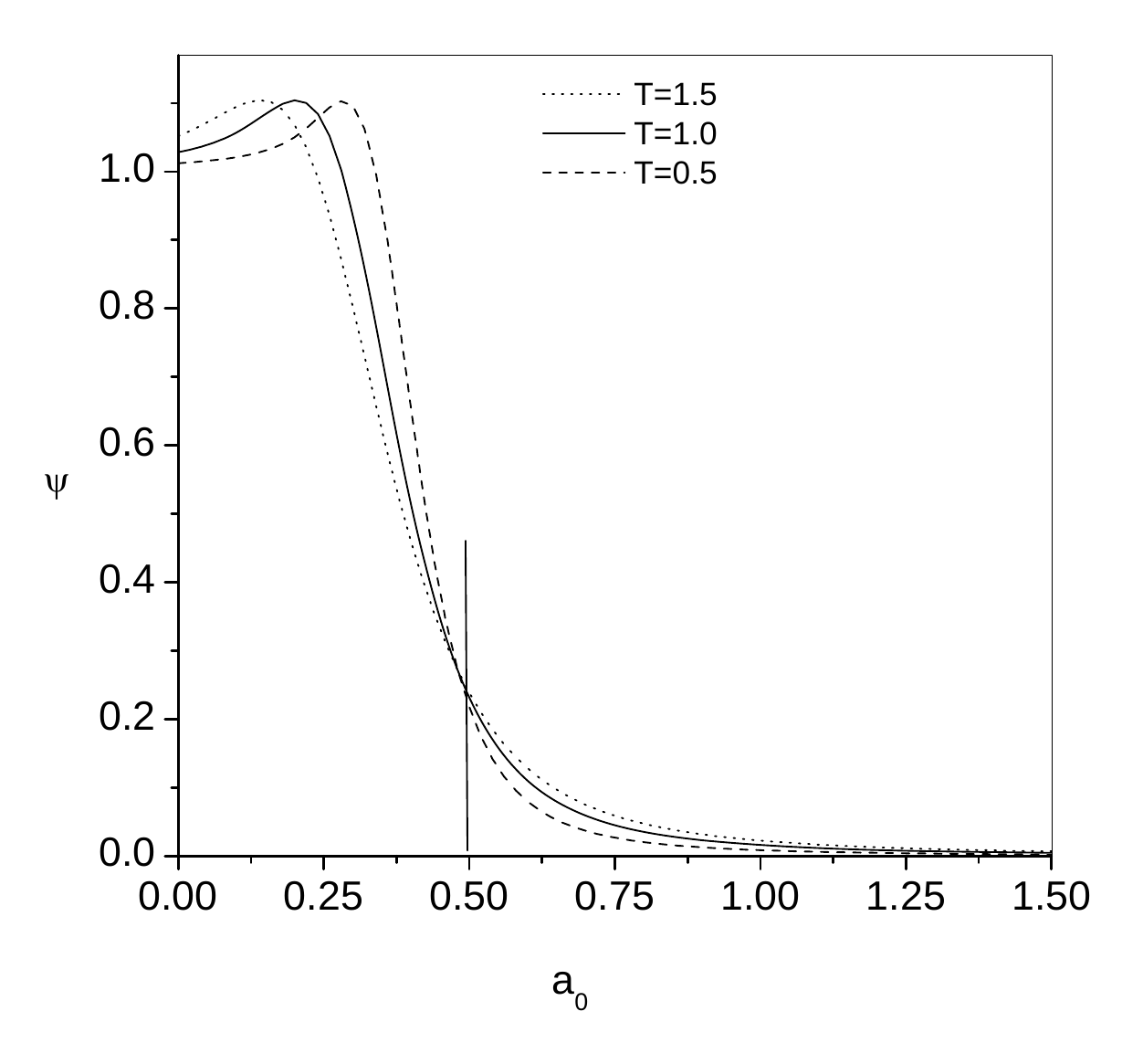}
\caption[]{Response function vs $a_0$ plot at different temperature
for the parameter set $a=0.5,\;b=0.005,\;K_B=1$ (using Eq.(2.9)).}
\end{figure}
The response function gradually increases to a maximum followed by
a sharp fall at around $a_0=a$. This type of variation of response
function with $a_0$ illustrates two types of dynamical
transition\cite{sancho}. (i) The first one corresponds to a
transition from an \emph{ordered} state to a \emph{dis-ordered}
state for $\Delta V= k_BT$. The barrier height of the potential
($\Delta V$) is $(a-a_0)^2/4b$. So with increasing value of $a_0$
($a_0>0$) the barrier height gradually decreases. At
$a_0=a-2\sqrt{bk_B T}$, the barrier height is double to average
kinetic energy of the particle. The response function($\Psi$) at
this point shows a maximum, presumably indicating the existence of
a disordering transition. The existence of the transition point can
be physically explained as follows: Whenever the value of average
kinetic energy crosses half of the barrier height the particles
start moving more randomly between two wells in a way as if they
(relatively large number of the particles) feel no barrier between
them. This is due to the fact that in this temperature domain the
distribution of kinetic energy gains a broad range. Here lies a
valid reason for existence a relation between barrier height and
kinetic energy in the transition region. As expected with the
increase in temperature the transition point shifts towards the
origin. (ii) The second type of transition is due to a change of
distribution function as reflected in Fig.3 by a very sharp
change(bifurcation) in the response function at $a_0=a$. This
transition is intrinsically different in nature from the previous
one which can be controlled by adjusting the temperature of the
system. In the later case, however, the temperature has no
significant influence.
\section{Jarzynski relation and work distribution for
 transition in nonequilibrium system}
For infinitely slow switching($\lambda=0$ to $\lambda=1$) of the
potential parameter(Eq.(\ref{1.3})), the system remains in
quasistatic equilibrium with the reservoir throughout the switching
process and the total work performed on the system will be equal the
Helmholtz free energy difference between initial and final state
states ($W=\Delta F$). If the switching process occurs with a finite
rate, the total work spent in changing the state of a system will
depend on the microscopic initial conditions of the system and the
reservoir and obey the following inequality
\begin{eqnarray}
\langle W \rangle >\Delta F\nonumber
\end{eqnarray}
In this case, the time evolution of the system of interest as
described earlier(Eq.(\ref{1.1a},b)) is governed by a stochastic
phase space trajectory, which depends on the externally imposed time
dependence of the switching parameter $\lambda (t)$. For the present
purpose we consider a constant switching rate, $\dot{\lambda}=1/t_f$
($t_f$ is the final time starting from $t=0$). With this time
dependence of switching parameter the total work performed along one
particular trajectory ($z(t)$) up to time $t$ is given by
\begin{eqnarray}
W[z(t),t]=\int_0^t dt\;\left(\frac{1}{t_f}\right) \frac{\partial
V}{\partial \lambda}(z(t))\label{1.8}
\end{eqnarray}
The equation for time evolution for work is given by
\begin{eqnarray}
\dot{W}=\frac{1}{t_f} \frac{\partial V}{\partial
\lambda}(z(t))\label{1.9}
\end{eqnarray}
Although the equation of motion for $W$ does not have an explicit
dependence on noise it is stochastic through its dependence on phase
space variables.
\begin{figure}[htp]
\centering\includegraphics[width=8cm,angle=0,clip]{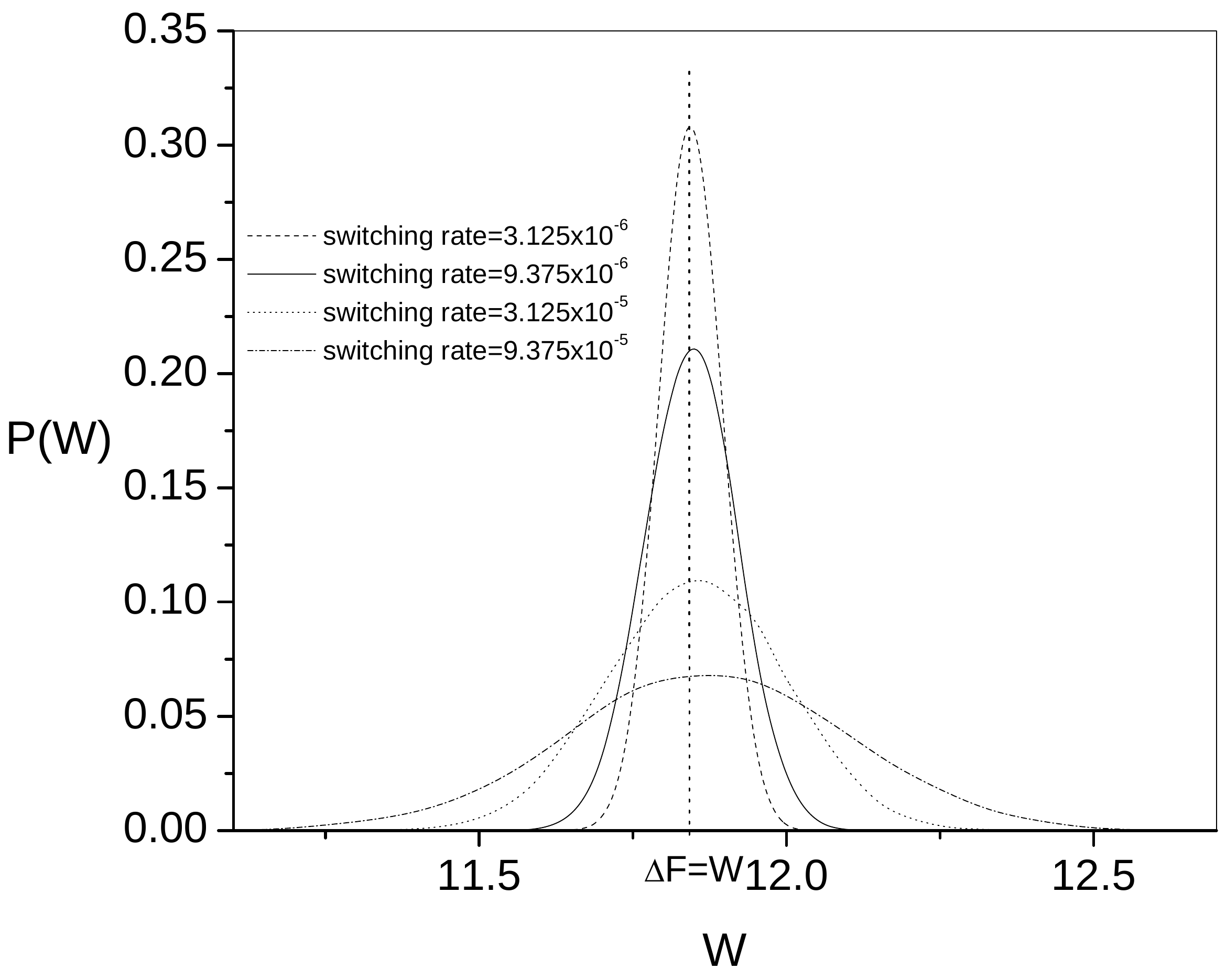}
\caption[]{Plot of work distribution function for different values
of switching rate for the parameter set $a=0.5,\;b=0.005,\;
T=1.0,\;K_B=1\;{\rm and}\; \gamma=2.5$.}
\end{figure}
Before we proceed to analyze the essential features of dynamical
transition in terms of the work done we first numerically simulate
the stochastic coupled equations Eq(\ref{1.1a},b) along with the
equation for work (\ref{1.9}) simultaneously using standard Heun's
algorithm and calculate the distribution of work, average work,
higher moments of work.  This allows us to calculate the free energy
change from Jarzynski relation as a non-equilibrium estimate. We use
in our numerical simulation a slowly varying time dependent quantity
$\lambda(t)$, with switching rate $\dot{\lambda}\sim
10^{-5}-10^{-6}$. A very small time step($\Delta t$) of $0.01$ for
numerical integration has been used. For the initial conditions we
have assumed that at $t=0$ all the particles are in the potential
minimum at $x=\sqrt{a/b}$ with zero velocity. In our simulation, we
first allow the system to equilibrate with the reservoir to smooth
out the effects due to the influence of initial conditions and
transient processes. After the equilibration process we switch on
the parameter ($\lambda(t)$). In calculation of average and higher
moments the averaging is done over 20,000 trajectories. The
parametric dependence of stochastic trajectory takes care of energy
balance\cite{suzki} for Langevin dynamics when work is calculated in
terms of Eq.(3.1).
\begin{figure}[htp]
\centering\includegraphics[width=16cm,angle=0,clip]{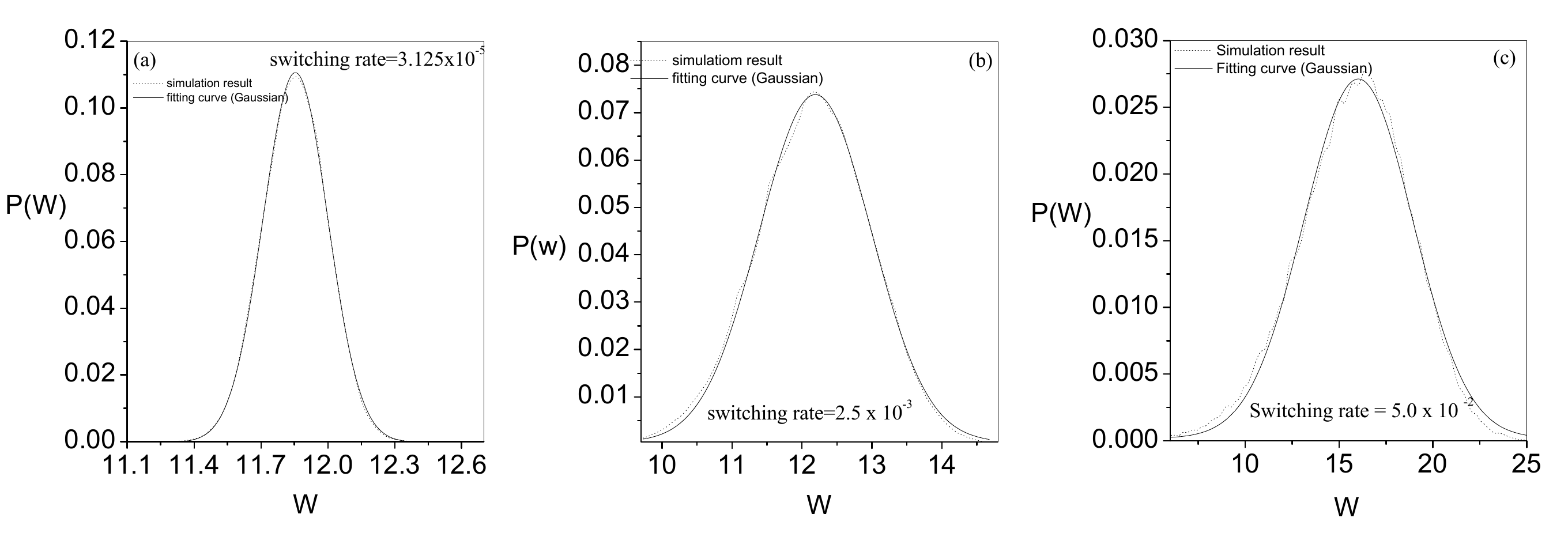}
\caption[]{ In this figure we have depicted work distribution
function for different switching rate with corresponding Gaussian
fitting curve. The parameter set: $a=0.5,\;b=0.005,\;
T=1.0,\;K_B=1\;{\rm and}\; \gamma=2.5$.}
\end{figure}

 We first calculate the work distribution, average
work and work fluctuation for different switching rate. This is
shown in Fig.4 and Table-I. As revealed by Fig.4 the distribution is
nearly  Gaussian for low switching rate centering around $\langle
W\rangle$, which is larger than free energy change( $\Delta F$) (The
vertical dotted line does not indicate the center of the
distribution but shown only to highlight the portion at which $W$
matches $ \Delta F$). With decreasing switching rate the center of
the distribution is shifted towards the line of free energy change,
$\Delta F=11.862536$ (average work done tends to be equal to the
free energy change along with the decrease of width of
distribution). The distribution function $P(W)$ tends to
$P(W)\rightarrow \delta(\langle W\rangle-\Delta F)$ . In Fig.5 we
have depicted the work distribution function along with
corresponding Gaussian fitting curve for relatively larger switching
rate ($\dot{\lambda} > 10^{-5}$). From Fig.5 it is apparent that the
work distribution function is not a Gaussian for arbitrary switching
rate. One observes that the higher order cumulants are important.
(The distribution will be a Gaussian only if the switching speed is
very slow for which the higher order cumulants effectively zero). We
present Table-I depicting the variation of average work and width of
the work distribution with switching rate.
\begin{tiny}
\begin{table}
\caption[] {The table presents nonequilibrium estimates of average
work and width of work distribution for different values of
switching rate for $a=0.5,\;b=0.005,\; T=1.0,\;K_B=1\;{\rm and}\;
\gamma=2.5$.}
\begin{tabular}{|c|c|c|}
  \hline
  Switching rate & $\langle W\rangle $ &
   \begin{tabular}{c}
  $\Delta W ^2$ \\($=\langle W^2\rangle-\langle W\rangle^2$)\\
   \end{tabular}\\

  \hline
 \begin{tabular}{c}
 $9.375\times 10^{-5}$\\
 \hline
 $3.125\times 10^{-5}$\\
 \hline
 $9.375\times 10^{-6}$\\
 \hline
 $3.125\times 10^{-6}$\\
 \end{tabular}
 & \begin{tabular}{c}
               11.889384\\
               \hline
               11.878023\\
               \hline
               11.870123\\
               \hline
               11.863661\\
             \end{tabular}
   & \begin{tabular}{c}
               0.049791\\
               \hline
               0.032513\\
               \hline
               0.017320\\
               \hline
               0.002563\\
             \end{tabular}\\
\hline
\end{tabular}
\end{table}
\end{tiny}
 To verify the Jarzynski relation we numerically estimate the
quantity $\langle e^{- W/k_BT}\rangle$. For a comparison of this
result with the analytically calculated free energy we present a
data Table-II for different values of $a_0$. The calculated free
energy change using Jarzynski relation
\begin{eqnarray}
\Delta F= -k_BT\ln{\left(\langle e^{- W/k_BT}\rangle
\right)}\label{2.3}
\end{eqnarray}
 matches almost exactly (absolute difference is less then $0.01$ $\%$).
\begin{tiny}
\begin{table}
\caption[] {The table presents a comparison of free energy change
calculated analytically (equilibrium estimate) and numerically using
Jarzynski relation (non-equilibrium estimate) for different values
of $a_0$. The other parameter set for this data table are
$a=0.5,\;b=0.005,\; T=1.5,\;K_B=1,\; \gamma=2.5$, and Switching rate
$ = 5 \times 10^{-6}$.}
\begin{tabular}{|c|c|c|c|c|c|}
  \hline
  $a_0$ & \begin{tabular}{c}
  $\Delta F$\\
  (using Eq.(2.8),\\equilibrium \\estimate ) \\\end{tabular}& $\langle W\rangle $
  &$\Delta W ^2$ & \begin{tabular}{c}
  $\Delta F$ (using Jarzynski\\ relation Eq.(\ref{2.3})\\non-equilibrium \\estimate)
\\\end{tabular}  & \begin{tabular}{c} $\Delta F=\langle W\rangle-
\frac{1}{2}\frac{\Delta W ^2}{k_BT}$\\
(fluctuation-dissipation\\estimate)\end{tabular}\\
  \hline
 \begin{tabular}{c}
 0.46\\
 \hline
 0.47\\
 \hline
 0.48\\
 \hline
 0.49\\
 \hline
 0.50\\
 \hline
 0.51\\
 \hline
 0.52\\
 \hline
 0.53\\
 \hline
 0.54\\
 \end{tabular}
 & \begin{tabular}{c}
               11.452822\\
               \hline
               11.522062\\
               \hline
               11.58799\\
               \hline
               11.650821\\
               \hline
               11.710751\\
               \hline
               11.767964\\
               \hline
               11.822633\\
               \hline
               11.874918\\
               \hline
               11.924969\\
             \end{tabular}
   & \begin{tabular}{c}
               11.456438\\
               \hline
               11.525216\\
               \hline
               11.592910\\
               \hline
               11.655346 \\
               \hline
               11.714050 \\
               \hline
               11.772241\\
               \hline
               11.827798\\
               \hline
               11.878662\\
               \hline
               11.929216 \\
             \end{tabular} & \begin{tabular}{c}
               0.011571\\
               \hline
               0.0119362\\
               \hline
               0.012375\\
               \hline
               0.012639\\
               \hline
               0.012955\\
               \hline
               0.013103\\
               \hline
               0.013666\\
               \hline
               0.014022\\
               \hline
               0.014020\\
             \end{tabular} & \begin{tabular}{c}
                11.452579\\
               \hline
                11.521237\\
               \hline
                11.588780\\
               \hline
                11.651129\\
               \hline
                11.709732\\
               \hline
                11.767872\\
               \hline
                11.823242\\
               \hline
                11.873984\\
               \hline
               11.924543\\
             \end{tabular}  & \begin{tabular}{c}
               11.452581 \\
               \hline
               11.521238\\
               \hline
               11.588785 \\
               \hline
               11.651133 \\
               \hline
               11.707931\\
               \hline
               11.767873 \\
               \hline
               11.823243 \\
               \hline
               11.873988\\
               \hline
               11.924542 \\
             \end{tabular}\\
\hline
\end{tabular}
\end{table}
\end{tiny}
As revealed by the data set of Table-II the following relation for
the free energy change with work holds very good as a
fluctuation-dissipation estimate
\begin{eqnarray}
\Delta F =\langle W \rangle-\frac{1}{2k_B T}\left\{\langle
W^2\rangle-\langle W\rangle^2\right\}\label{2.4}
\end{eqnarray}
the dissipative work being equal to the width of the distribution of
work,
\begin{eqnarray}
w_{diss}=\frac{1}{2k_B T}\left\{\langle W^2\rangle-\langle
W\rangle^2\right\}\label{2.5}
\end{eqnarray}
A final remark on Table-II may be in order. Since the work
fluctuations have been determined with the help of Langevin dynamics
with a Gaussian noise, fluctuation-dissipation estimate in Eq.(3.4)
matches well with free energy change. For a different protocol for
following the trajectory or for higher switching rate the deviation
from linear response is expected where non-Gaussian distributions of
work make their presence felt. The latter aspect is evident in
Fig.5.
\section{Study of dynamical transition by work fluctuation and response function}
We recall that a system out of equilibrium can not, in principle,
be described by free energy and therefore no free energy functional
or partition function methodology can be applied for
non-equilibrium system for classification of dynamical transition.
Guided by a close analogy with equilibrium phase transition in
Sec.II we have identified an appropriate response function $\Psi$
for description of dynamical transition. Analysis of Sec.III on the
other hand suggests that by virtue of Jarzynski relation one can
relate the free energy change (rather than free energy itself)
between two thermodynamic states to microscopic work fluctuation
for non-equilibrium paths connecting these states. The question is,
can we bypass the description based on free energy change as done
in Sec.II to compute the response function $\Psi$ directly from
work fluctuation and recover the features of non-equilibrium
dynamical transition. To this end we now proceed to calculate the
response function $\Psi$ in terms of the work performed using
Eq.(\ref{2.4}) and Eq.(\ref{2.5}) as follows;
\begin{eqnarray}
\Psi=-\frac{\partial ^2\Delta F}{\partial a_0^2} =-\frac{\partial
^2\langle W\rangle}{\partial a_0^2}+\frac{\partial ^2
w_{diss}}{\partial a_0^2}\label{3.1}
\end{eqnarray}
From Eq.(\ref{1.8}) and (\ref{1.3}) the work performed for the
switching process along a particular trajectory is given by
\begin{eqnarray}
W[z(t)]=\int_0^{t_f} dt\;\left(\frac{1}{t_f}\right) \frac{1}{2}a_0
\;x^2\label{3.2}
\end{eqnarray}
As revealed by the above expression the work performed is a linear
function of $a_0$ only if the process $x(t)$ is independent of
control parameter($a_0$). In this situation we have
\begin{eqnarray}
\frac{\partial \langle W\rangle}{\partial a_0}=\frac{\langle
W\rangle}{a_0}\;;\;\;\;\;\;\;\;\frac{\partial^2 \langle W
\rangle}{\partial a_0^2} =0\label{3.3}
\end{eqnarray}
With the help of the Eq.(\ref{3.3}) the response function can be
expressed in a simplified form as
\begin{eqnarray}
\Psi=\frac{1}{a_0^2}\frac{\langle W^2 \rangle-\langle W
\rangle^2}{2k_BT}\label{3.4}
\end{eqnarray}
\begin{figure}
\centering\includegraphics[width=8cm,angle=0,clip]{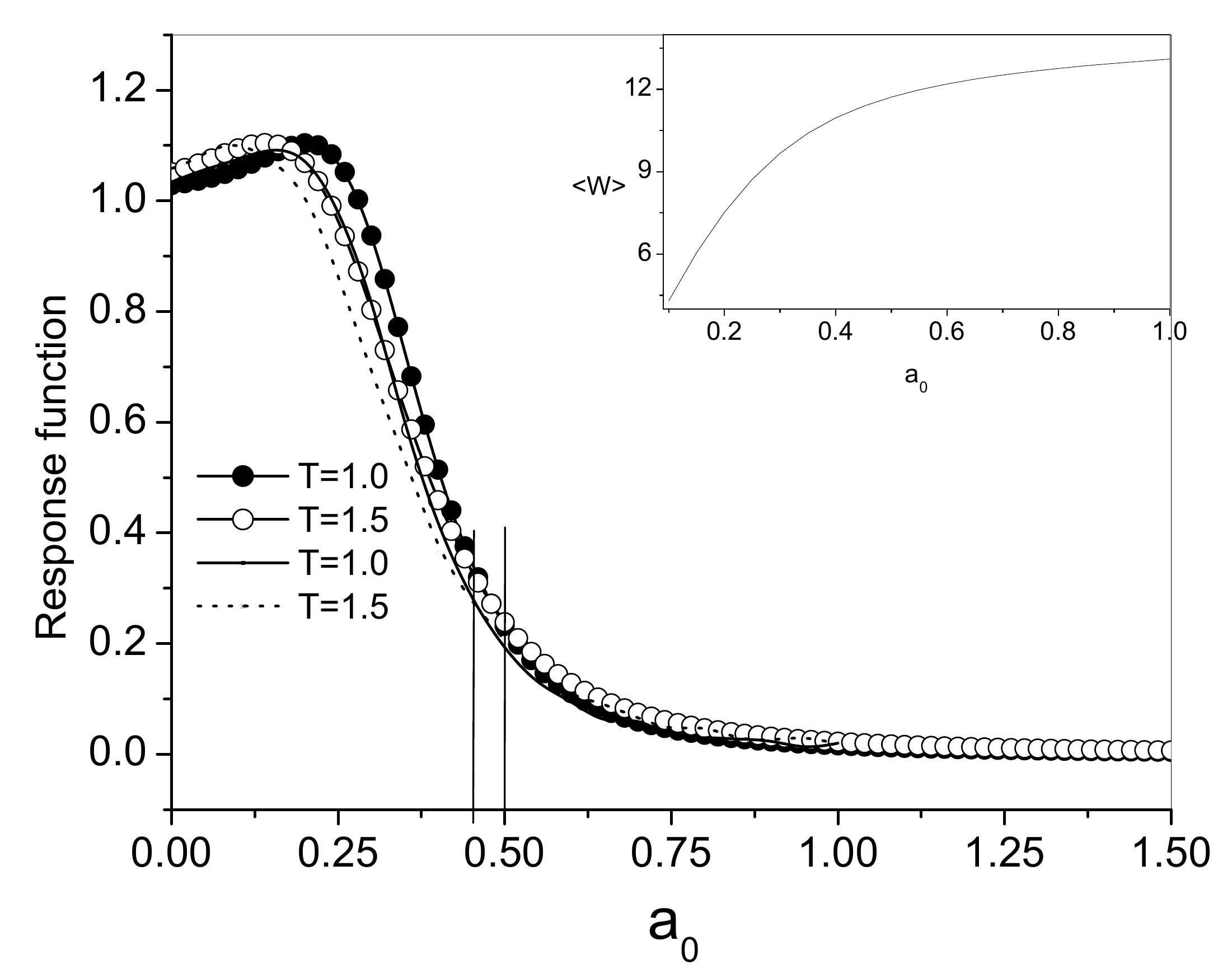}
\caption[]{The solid and dotted lines present the variation of
response function without dissipative work  as a function of $a_0$
for different values of temperature. The lines containing hollow and
solid circles present the variation of response function with
inclusion of dissipative work as a function of $a_0$ for different
values of temperature. The parameter set used
$a=0.5,\;b=0.005,\;\gamma=2.5\;K_B=1$ . The inset plot presents the
variation of work as a function $a_0$ for the same parameter set as
in the main figure but for $T=1.0$, Switching rate $= 10 ^{-5}$.}
\end{figure}
The response function is thus related to dissipative work. The above
expression for response function has a close similarity to other
linear response functions, as for example, heat capacity
$C_v=\frac{\langle E^2 \rangle-\langle E \rangle^2}{2k_BT}$,
magnetic susceptibility $\chi=\frac{\langle M^2 \rangle-\langle M
\rangle^2}{2k_BT}$, where $E$ and $M$ denote internal energy and
magnetization, respectively. A difference between linear response
functions and $\Psi$, however, is noteworthy$^{11}$. Since $E$ and
$M$ are well-defined equilibrium properties of a macroscopic system
in the thermodynamic limit while  $W$ by its very nature corresponds
a quantity defined for  non-equilibrium paths, heat capacity and
susceptibility are typical static response function in contrast to
the non-equilibrium response $\Psi$.
\begin{figure}
{\includegraphics[width=5cm,angle=-0]{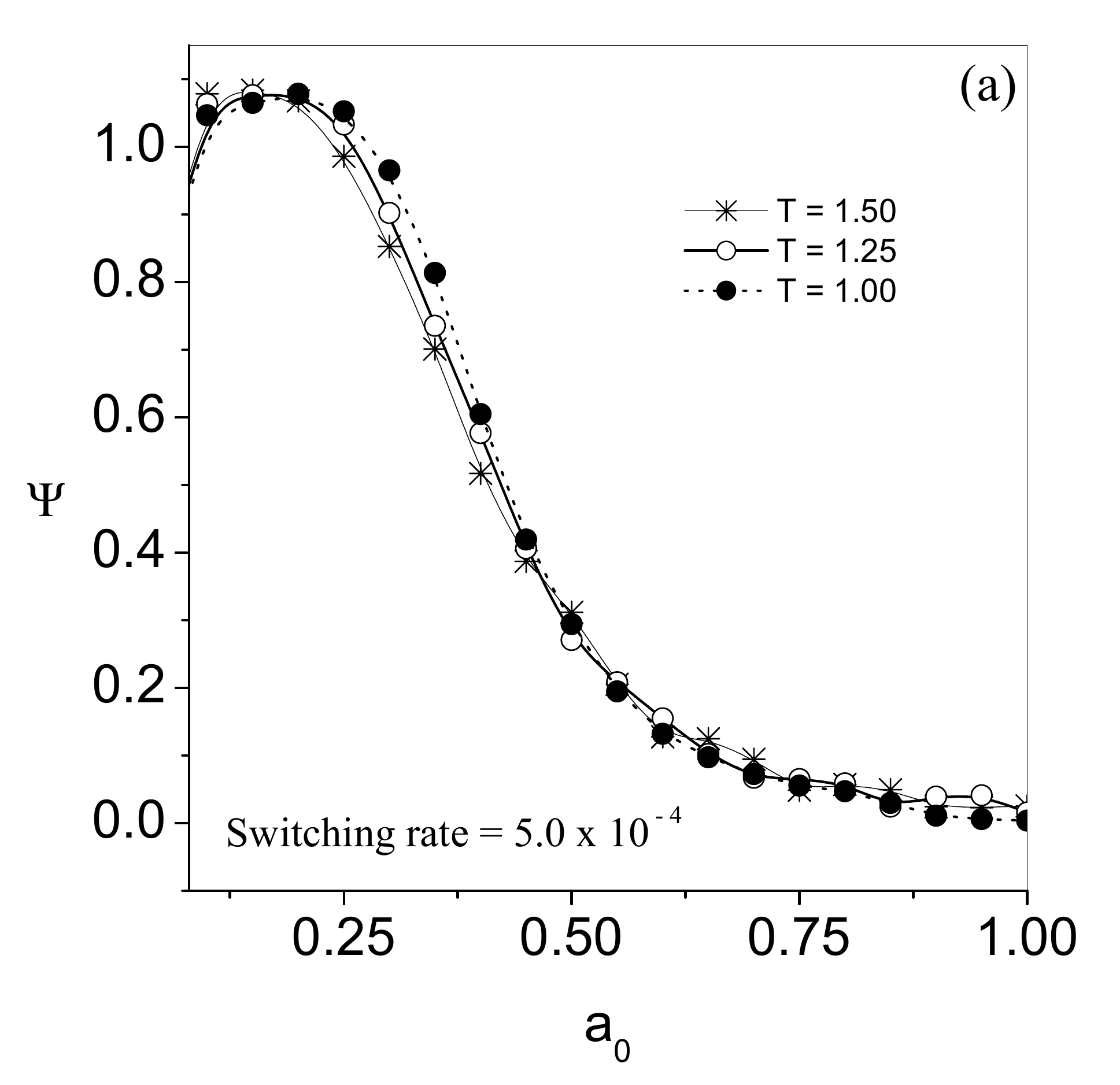}}\quad
{\includegraphics[width=5cm,angle=-0]{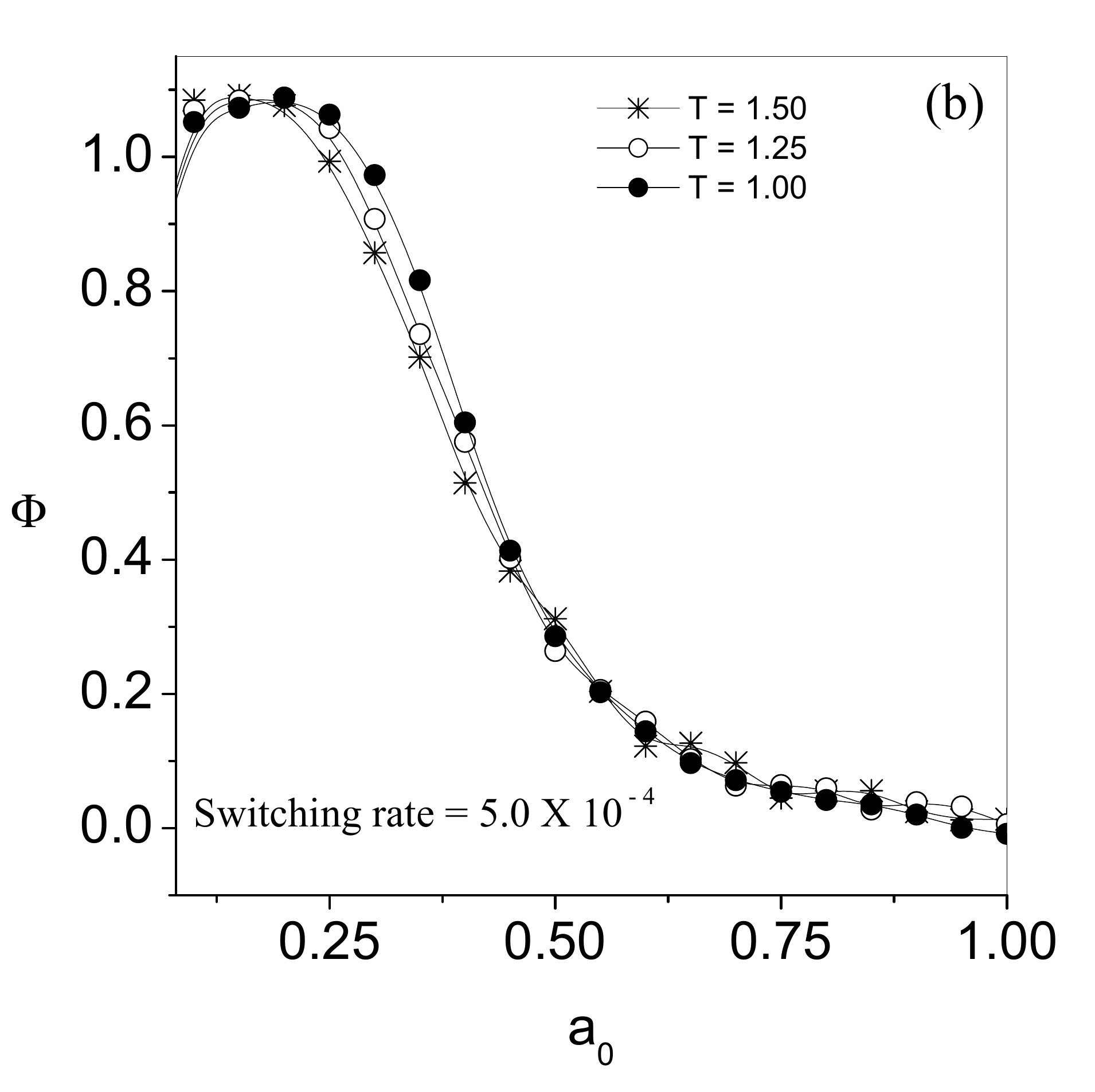}}\\
{\includegraphics[width=5cm,angle=-0]{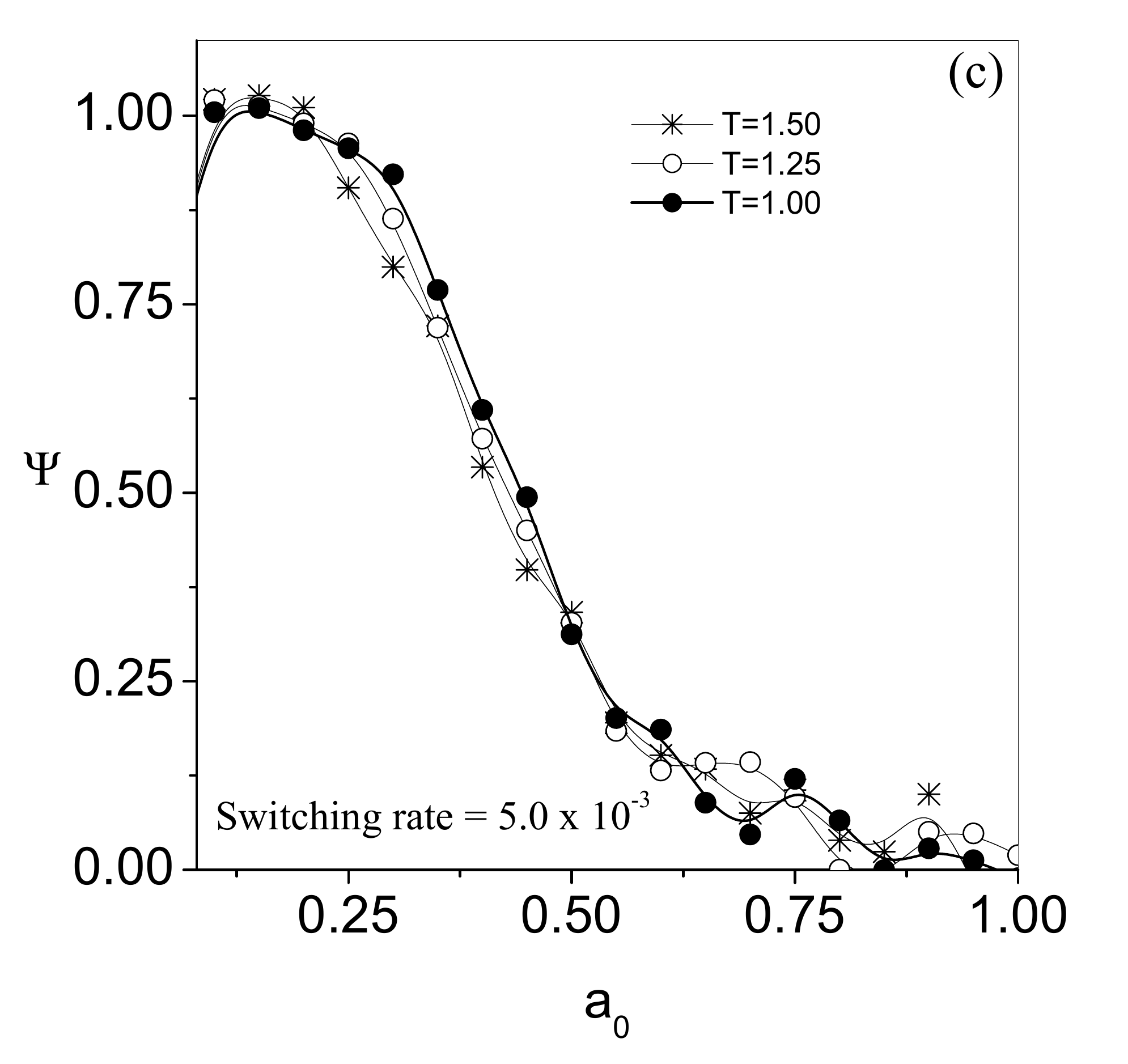}}\quad
{\includegraphics[width=4.7cm,angle=-0]{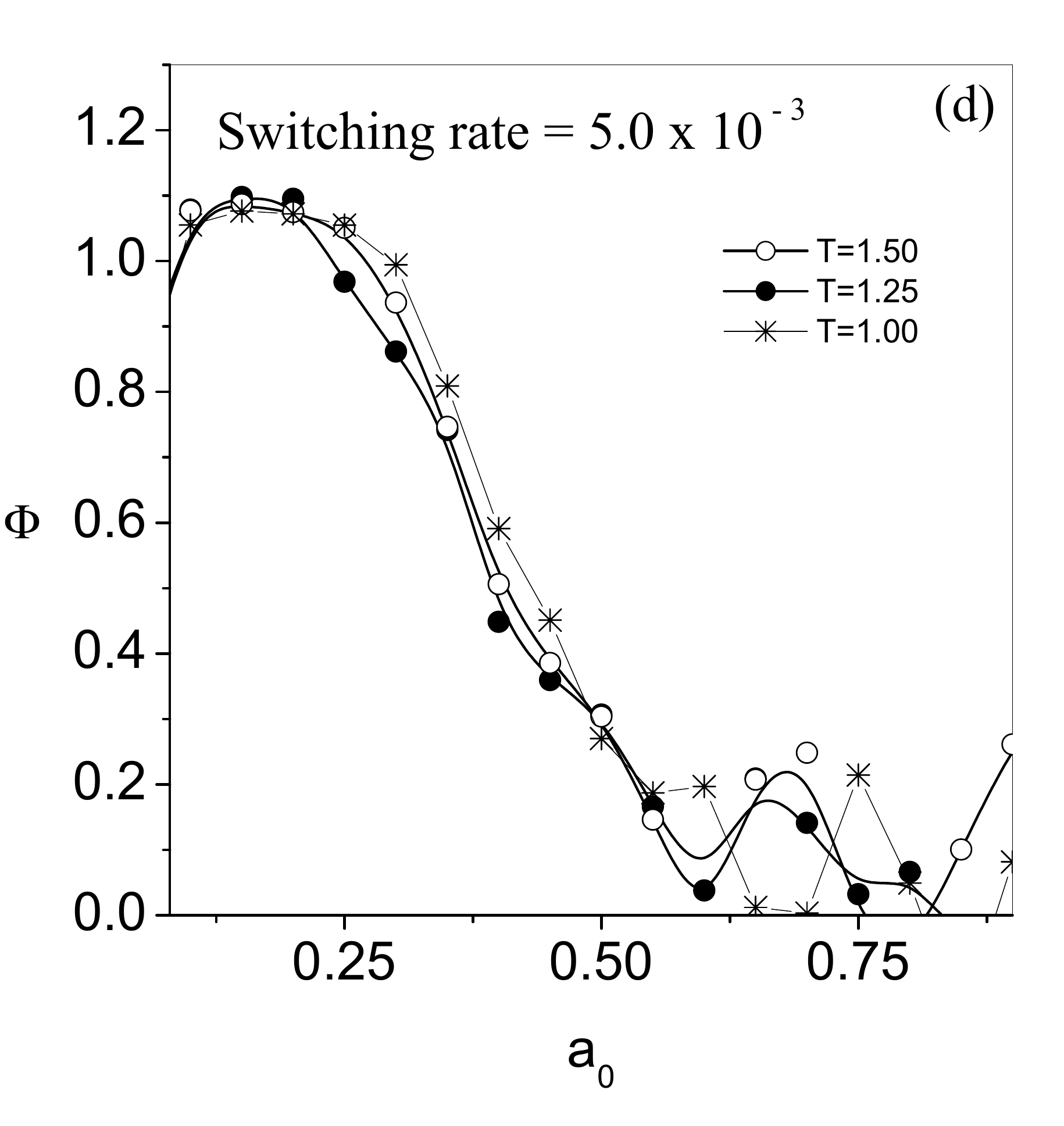}}\\
{\includegraphics[width=5cm,angle=-0]{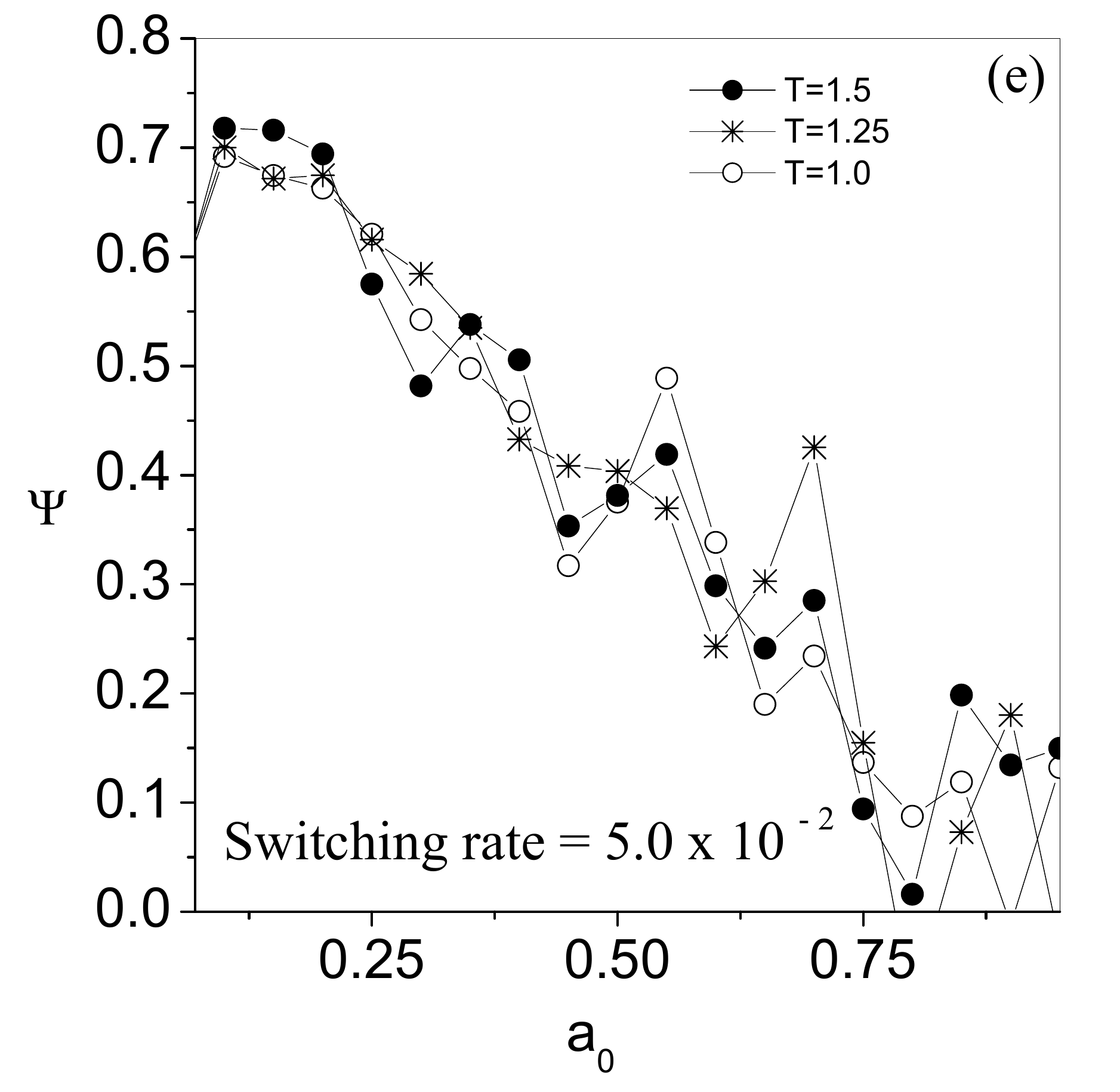}}\quad
{\includegraphics[width=5cm,angle=-0]{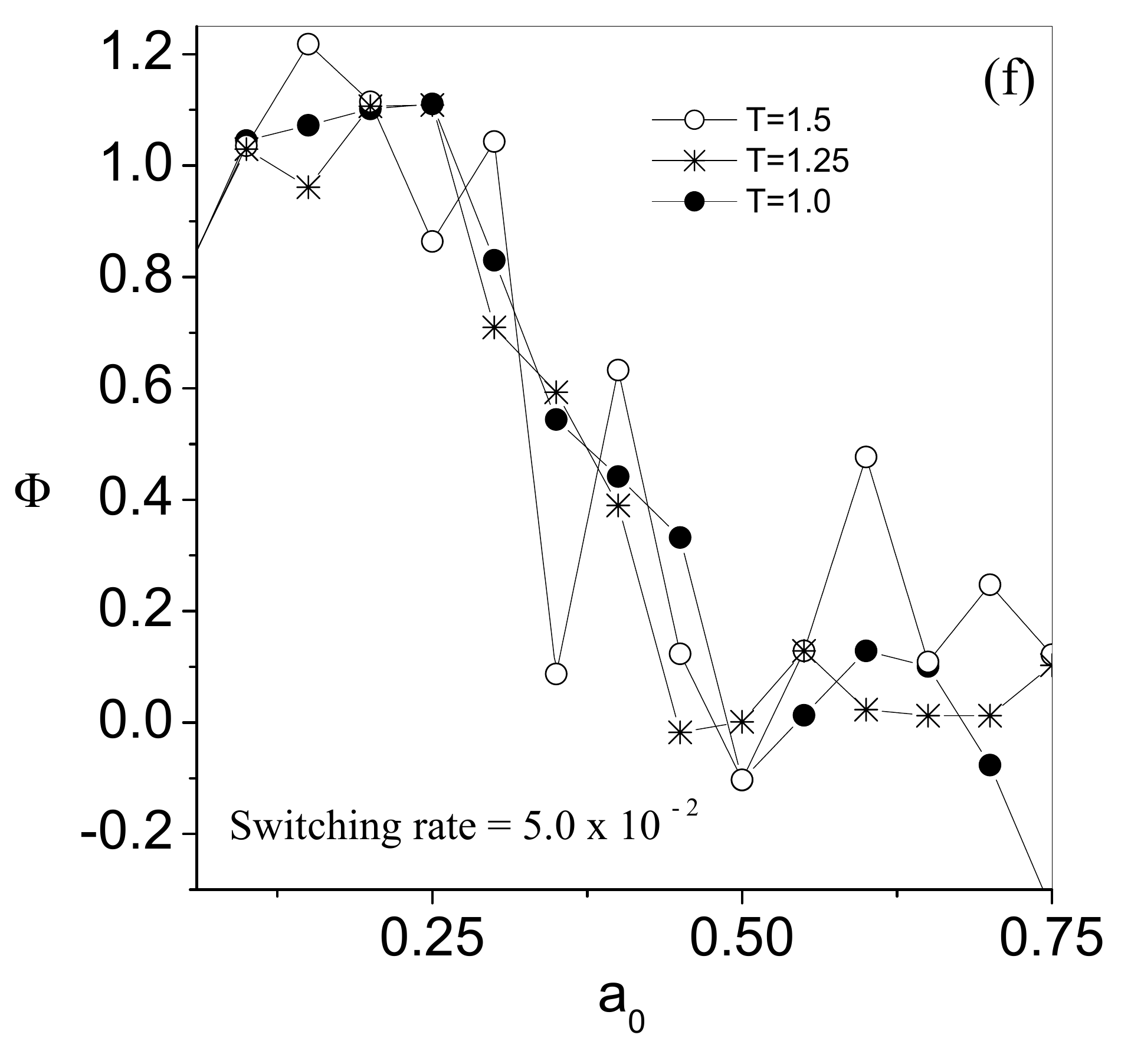}} \caption[]{ The
subfigures (a, c, e) present the variation of response function
$\Psi$ with $a_0$ and the subfigures (b, d, f) present the variation
of $\Phi = - \frac{\partial^2 \langle W \rangle}{\partial a_0 ^2}$
with $a_0$ depicting dynamical transition for different switching
rate. The parameter set used $a=0.5,\;b=0.005,\;\gamma=2.5\;K_B=1$.}
\end{figure}
In the present problem the work, however, is not a linear function
of $a_0$. This is due to the fact that $a_0$ bears an implicit
dependence on $x(t)$. The non-equilibrium estimation of $\langle W
\rangle$ shows that,in general, it is a nonlinear function of $a_0$,
as depicted in the inset of plot Fig.6. $\langle W \rangle$ behaves
linearly only for small values of $a_0$. The second derivative of
the average work also predicts two transition points as discussed in
Sec.II. But the transition points are slightly shifted from original
position of the transition point (this is depicted in Fig.6). It
should however be emphasized that although the second order
variation of average work predicts only the approximate transition
point, the inclusion of the contribution due to dissipative work
gives an accurate estimate of the transition points as shown in
Fig.6 (the plots containing solid and hollow circle).

In order to check the dependence of response function on switching
rate we present the variation of response function as a function of
$a_0$ for different switching rate in the Fig.7 (a,c,e). One
observes that fluctuations become larger for higher switching rate.
It may, however, be checked that the nature of the variation of
response function with $a_0$ remains same. Although the nature of
variation of the response function indicates two types of dynamical
transition for arbitrary switching rate, but it is difficult to
point out the transition points for higher switching rate due to
fluctuating nature of $\Psi$. Moreover in order to figure out the
relative contribution of second derivative of average work, \emph{i.
e.}, $-\frac{\partial^2 \langle W \rangle}{\partial a_0 ^2}$ and
that of dissipative work $-\frac{\partial^2  W_{diss}}{\partial a_0
^2}$ in response function we present the subfigures of Fig.7 as a
pair (a,b), (c,d) and (e,f). While Figs.7(a,c,e) represent variation
of response function including dissipative work, Figs.7(b,d,f)
represent the cases without dissipative work. From these figures it
is clear that one can identify the dynamical transitions by
calculating only second derivative of average work, \emph{i. e.},
$\frac{\partial^2 \langle W \rangle}{\partial a_0 ^2}$. Thus
 the nature of the variation of response functions derived from work
fluctuations are practically independent of any pulling speed which
is a correct reflection of Fig.3. The response function(4.1) or
(4.4) is completely determined by the stochastic equations (2.1) and
(3.1) of the dynamical system independent of free energy description
of the system in the thermodynamic limit.

\section{Conclusion}
Analysis of transition between two equilibrium states is
traditionally based on free energy change in a system with respect
to a relevant parameter and linear response function is related to
fluctuation of the order parameter around equilibrium. In view of
the fact that free energy remains undefined for non-equilibrium
systems, any analysis of response function on the basis of similar
argument is untenable. However as Jarzynski equality is related to
free energy change with work fluctuations during the passage of the
system over many non-equilibrium paths when the parameter of the
potential is varied, it is possible to look for the signature of
dynamical transition in the behaviour of work fluctuations. The key
quantity is the response function, which exhibits a characteristic
behaviour of the system itself with no limitation imposed by
near-equilibrium condition. Based on a model system we have examined
two types of transition, e., g., order-disorder type and a
bifurcation type which can be differentiated by their thermal
behaviour around the transition points. As the Jarzynski relation
and the related fluctuation theorems are valid even far from
equilibrium situations, the response functions derived from work
fluctuations, we believe, have a wider range of applicability, i.,
e., beyond linear regime and the results obtained for this simple
system may be extended to explore more complex issues.

\acknowledgments Thanks are due to the Council of Scientific and
industrial research, Govt. of India, for partial financial support.

\newpage


\begin{thebibliography}{99}
\bibitem{jar1} C. Jarzynski, Phys. Rev. Lett. {\bf 78}, 2690 (1997); J. Stat. Mech.:
Theory Exp., {\bf P09005} (2004).


\bibitem{jar2} C. Jarzynski, Phys. Rev. E {\bf 56}, 5018 (1997).

\bibitem{crooks} G. E. Crooks, Phys. Rev. E {\bf 60}, 2721 (1999).

\bibitem{wang} G. M. Wang, E. M. Sevick, E. Mittag, D. J. Searles, and D. J.
Evans, Phys. Rev. Lett. {\bf 89}, 050601 (2002).

\bibitem{Carbe} D. M. Carberry,
J. C. Reid, G. M. Wang, E. M. Sevick, D. J. Searles, and D. J.
Evans, Phys. Rev. Lett. {\bf 92}, 140601 (2004).

\bibitem{Liphardt} J. Liphardt, S.
Dumont, S. B. Smith, I. Tinoco and C. Bustamente, Science {\bf 296},
1832 (2002).

\bibitem{hummer} G. Hummer and A. Szabo, PNAS {\bf 98}, 3658 (2001).

\bibitem{bussi} G. Bussi, A. Laio, and M. Parrinello, Phys. Rev. Letts {\bf 96}, 090601
(2006).

\bibitem{hatano} T. Hatano, Phys. Rev. E, {\bf 60} R5017 (1999).

\bibitem{speck} T. Speck and U. Seifert, Phys. Rev. E {\bf 70}, 066112
(2004).

\bibitem{rito} F. Ritort; Poincare Seminar, {\bf 2}, 193 (2003) and
references therein.

\bibitem{sancho} J. Garc\'{i}a-Ojalvo, and J.M. Sancho,
\emph{Noise in Spatially Extended Systems} (Springer-Verlag, New
York, 1999).

\bibitem{pkg2} P. K. Ghosh, D.
Barik and D. S. Ray, Phys. Lett. A {\bf 342} 12 (2005).

\bibitem{ff5} M. Borromeo and F. Marchesoni, Europhys Lett {\bf 68}, 784 (2004);
M. Marchi et al. Phys. Rev. E {\bf 54}, 3479 (1996).

\bibitem{suzki} D. Suzuki et al., Phys. Rev. E {\bf 68}, 021906 (2003).



\end{thebibliography}
\end{document}